\documentclass[prd,nobibnotes,nofootinbib,showpacs,12pt]{revtex4}
\pdfoutput=1
\usepackage[latin1]{inputenc}
\usepackage{graphicx}
\usepackage{amsmath}

\numberwithin{equation}{section}
\usepackage[mmddyyyy]{datetime}
\newdate{date}{30}{11}{2016}

\begin{document}
\title{Resolving-Power Quantization.}
\author{Herbert Neuberger}
\email{herbert.neuberger@gmail.com}
\affiliation{Department of Physics and Astronomy, Rutgers University,\\ 
Piscataway, NJ 08854, U.S.A} 

\begin{abstract}
Starting with a general discussion, a program is sketched for 
a quantization based on dilations. This resolving-power quantization is simplest for scalar field theories. 
The hope is to find a way to relax the requirement of locality so that the necessity to fine tune mass parameters is eliminated while universality is still preserved.
\end{abstract}
\date{\displaydate{date}}
\maketitle


%
%


\section{Introduction}

In the Path Integral language, quantization amounts to a segmentation of the set of integration variables which orders 
them into subsets of identical type.

The integration variables are functions of space-time. It is often possible to restrict to Euclidean signature and this often maps the problem into one of Probability Theory on function spaces. I am focusing on this simplest of situations. 

A foliation of space-time induces a decomposition of the set of functions on space-time. 
Flat space-time $R^d$ can be foliated into $R^{d-1}$ slices, producing the standard equal-time quantization, or into $S^{d-1}$ slices, yielding radial quantization. Depending on the vector field perpendicular to the slices, the symmetry group of flat $R^d$ gets decomposed in different ways. Standard quantization singles out rotations and translation in $R^{d-1}$, while radial quantization singles out the rotation group of $R^d$~\cite{radqu}. There is no rule saying that any segmentation of the integration variables of the Path Integral has to be induced by a foliation of space-time.

In Nature, (uniform) scalings (similarity transformations or dilations) of $R^d$ seem to play a central role. Radial quantization provides a convenient framework to study dilations, but then translations all but disappear. 

K. Wilson, in two related papers in the early seventies laid the foundations of the ``Wilsonian Renormalization Group'' (RG)~\cite{wilson1, wilson2}. The second paper employed a phase-space cell analysis to make it concrete how the RG would work in a semi-quantitative manner. The phase-space cell analysis produced a new type of segmentation of the Path Integral. One was supposed to integrate sequentially over slices of fields separated by resolution level. To a Harmonic Analyst this probably points to Littlewood-Paley theory~\cite{litpal}. 

A far reaching consequence of the RG is that sometimes a non-empty finite set of parameters defining the microscopic theory need to be ``fine tuned'' in order to produce a useful macroscopic theory. The flip side of the coin is that the resulting macroscopic theory is largely independent of almost all the parameters defining the microscopic theory. 

This ``universality'' makes Field Theory a useful framework to describe Nature because phenomena at some level of resolving-power can be efficiently described without knowledge of the Physics laws at much higher resolutions. The price is, some of the time, that fine-tuning is required. 
My hope is that we can keep the good part of the RG construction, universality, but avoid paying the price of fine-tuning. 

Maybe we shall manage to first break through the barriers blocking attempts to synthesize Quantum Mechanics and General Relativity. We very much would like to achieve this synthesis. 
It is not proven that the single possible solution to the fine tuning problem of the Standard Model, including the $\sim 125\; GeV$ Higgs particle, has to be the one chosen by Nature, which in one form or other must include a ``Quantum Gravity''. This is my 
basic premise. 

An analogue is the much simpler problem of chirality. We can deal with it in Lattice Field Theory, but the solution chosen by Nature could be very different.  

Therefore, I decided to search for some untraditional solution to the fine tuning problem in simplified (but non-trivial) models. A careful walk through the steps taken by Wilson might lead to the discovery of one assumption that can be relaxed so as to avoid fine tuning at an acceptable cost. An obvious candidate is space-time locality. 

In the chiral analogy one could make progress by abandoning the (implicit) assumption that the number of fermion fields at the microscopic level is finite~\cite{chiral}. The infinite number of microscopic fields causes an indeterminacy in the gauge field functional resulting from the integration over all Fermion fields, and this indeterminacy makes room for the standard chiral anomaly to defeat any attempt to eliminate it in a way continuous in the gauge fields. For vector-like theories there is no indeterminacy to start with. 
For a theory where the chiral gauge anomalies do not cancel out, 
there is a fundamental obstruction to resolve the phase indeterminacy. When anomalies do cancel, this obstruction disappears.  Explicit elegant constructions in the latter case are still missing and some experts regard this as a cloud that conceptually needs dispersing. The integration over 
an infinite microscopic number of Fermion fields produces a violation of some strict form of locality, but a weaker form holds, and is sufficient for preserving the universality of the resulting continuum Field Theory. 

In short, take the simplest example of a chiral vector-like theory. Early on, Lattice Field theorists concluded that in the standard framework quarks of exactly zero mass required fine tuning. A minor relaxation of the standard 
framework eliminated this fine-tuning. To be sure, in this case, Continuum Effective Field theorists never had a fine tuning problem; actually, supersymmetry is a solution to fine tuning of scalar masses using the absence of fine tuning of fermion masses. 
But, the problems come back when one demands a fully defined Path Integral. 

I hope for some form of relaxation of the implicit assumption of only a finite number of integration variables in the execution of a (typically dyadic -- that is, involving powers of 2) single step of RG which could produce an automatic turn-off of the relevant operators in the vicinity of the fixed point of interest. Such a relaxation may be forced by some lack of locality in the microscopic action.

\section{Radial Quantization}

In Radial Quantization (RQ), after changing variables from $r=\sqrt{x_\mu x_\mu}$ to $\log (r)$ space-time becomes a cylinder $S^{d-1}\times R$ with dilations corresponding to translation in $R$. It is then possible to regulate the theory while preserving
an infinite abelian group of translations along $R$ and therefore a continuum theory would be invariant under dilations centered at
$x_\mu=0$. However, translations in $R^d$ cannot be preserved by the regularization. Without translations, the affine nature of dilations is lost and the very term ``dilation'' no longer implies the group theoretical role it suggests~\cite{radqu1}. 

My work on regularized RQ consisted of replacing the $R$-factor of the space-time cylinder by an equally spaced lattice and the $S^{d-1}$ factor by a refinable mesh of icosahedral symmetry at $d=3$.\footnote{One cannot approximate 3d rotational invariance in the same manner as $R^3$ translational invariance (e.g. by a cubic lattice) for a deep reason, related to the Banach-Tarski paradox. The source of the paradox is that  the free group of two generators, $F_2$, is isomorphic to a discrete subgroup of $O(3)$. This 
obviously extends to higher dimensions. This fundamental difference between $d\le 2$ and $d\ge 3$ advises caution in later discussions in the context of the $AdS_{d+1} /CFT_d$ correspondence and wavelets.}
 $R^3$ translational invariance imposes spectral regularities in the putative continuum limit of the transfer matrix in the $R$ direction. These constraints seem to be implementable without \underline{fine} tuning, by simple interpolation. I expected that as the mesh gets refined there will be more couplings that need to be adjusted (up to a point), but hoped that there always would be some comfortable ranges where the spectral constraints would be approximately satisfied. The refinement of the mesh was to be done by a 
``cubature'' strategy designed to make the action closer and closer to rotationally invariant in the sense that continuum action densities with components of angular momenta $l\le L$ 
would be exactly reproduced on the mesh with the integer $L$ increasing with the refinement. 

Using just a 12 vertex mesh a reasonable caricature of the low lying portion of the set of anomalous dimensions of the 3d Wilson-Fisher fixed point was obtained. Subsequent experimentation left unclear how the approach to continuum could be quantified and whether an infinite number of approximate adjustments would be needed to get to the continuum limit. Were that the case, whether fine-tuning one parameter is more of a blemish than an infinite number of coarse tunings is a question I am not sure 
is useful to address~\cite{radqu2}.

In one aspect RQ is very ambitious: it makes it easy to also preserve inversion $x_\mu \to x_\mu/x^2$. Inversion together with translations, rotations and dilations generates the entire conformal group. This means that the violation of translations is major. Restoring translations in the continuum limit would produce a conformal theory. Maybe RQ is too ambitious.

\section{AdS/CFT}

Another option for a different kind of quantization method might be offered by the famous $AdS_{d+1}/CFT_{d}$ correspondence~\cite{Natsuume}. 
I will stay in Euclidean signature and not attempt to define what this correspondence precisely is. It contains gravity in the $d+1$-dimensional bulk and has an ordinary conformal field theory on the $d$-dimensional boundary. One needs an $AdS$ asymptotic condition on the bulk. In this paper I make no attempt to address the Lorentz signature case. 

My first step is to get rid of gravity in the bulk by replacing the bulk by an infinite rigid lattice. I just give up on the most exciting parts of the conjectured equivalence. 
The lattice is the Cayley graph of an infinite discrete group. The infinite discrete group is a subgroup of the isometry group of $d+1$ dimensional Hyperbolic Space. The Poincar{\' e} construction identifies this group with the group of conformal transformations of one point compactified $d$-dimensional flat Euclidean space. This conformal group is the $d$-dimensional group of M{\" o}bius transformations~\cite{madrid, ratcliffe}.

The Cayley graph interpretation first selects one minimal set of generators for the discrete group. This set is finite. Next, the Identity element of the group is attached to one chosen site. The sites are connected by links, one for each generator.
We navigate from this center to other sites by using the link  generators. Each site of the infinite lattice is associated with one element of the discrete subgroup. The sites can be labelled in an order that is non-decreasing with the distance from the center. The distance is the minimal number of generators that constitutes a word expressing the group element  of the site. One can define shells of sites of equal distance from the center. Now we can truncate the graph by stopping after the completion of some shell. That shell is the boundary of the bulk which consists of all sites closer to the origin. The lattice looks like a Bethe--lattice, but crucially includes extra links which violate the Bethe--lattice tree structure. The counting of sites remains  similar to that of the Bethe-lattice, so the number of sites per shell increases exponentially. We are dealing with an ``expander graph''. Embedding the graph into Hyperbolic space places the Identity of the discrete group at some arbitrarily fixed point. 
All the links have equal length in the $AdS$ metric and therefore the lattice itself is homogeneous. We still need to choose an action. 

It is easy to imagine an action which is local in the bulk sense and includes the boundary of the truncated graph. Integrating the strictly bulk fields will induce an action on the boundary. The number of variables in the strict bulk is of the same order of magnitude as the number of sites on the boundary because of the exponential growth of the shells.
The induced action on the boundary does not look local, but we really don't know what this means. There is a distance between sites making up the boundary of the truncated lattice when seen in the infinite lattice. But that distance is not appropriate for the boundary. 

Truncating the Cayley graph destroys the group. To restore it in the continuum limit, we must take
some kind of distance of the truncation boundary from infinity to zero and adjust, if needed, the action. If the discrete group is restored, it likely will automatically ``flesh out'' to the continuum conformal group. 
The natural metric is the hyperbolic one and any boundary would be infinitely far from infinity. So, we need to choose some other metric in which the boundary in hyperbolic space can be made as close to infinity as one desires. In continuum one can reduce the ambiguity in this procedure to no more than an arbitrary conformal factor for a new metric on the boundary. Thus, the construction only produces a conformal class of boundary metrics. 

The fundamental trouble with this approach is that, even if we forget about the truncation and its elimination, we have an ill defined Statistical Physics problem because there is too much boundary. 
In the usual case, one can get directly at the infinite system limit (the thermodynamic limit) using conditional probabilities and imposing the 
Dobrushin-Lanford-Ruelle (DLR) equations. But, here there is no
unique Gibbs state (solution to the DLR equations): rather there is an infinity of them~\cite{sinai2}. So, even
if the truncation produced some continuum theory, another truncation might produce a different one. It seems that more needs to be understood about universality in this scheme, before one can make an attempt at a concrete construction.

There are two valuable lessons I draw from this. (i) The different dimensions of the continuum bulk and boundary are no mystery; they naturally can be discretized by meshes of comparable number of vertices. (ii) The ``extra dimension'' ought to correspond somehow to resolving-power on the boundary segmented out in the bulk. This points in the direction of wavelets. Before we go there, I turn to another pointer to wavelets, having to do with ``continuous smearing'' in the Path Integral formulation. Continuous smearing can be thought of as a stochastic version of gradient flow.

\section{Continuous Smearing}

Relaxing the finiteness of the number of local fields is an implicit part of any discussion of the $N=\infty$ limit of field theories whose symmetry groups are one of the $N-$groups. 

4d pure YM $SU(N)$ continuum gauge theory provides an example where scale dependence becomes singular at $N=\infty$. A singularity appears in certain operators and not in others; it reflects a dynamical physical property of the Field Theory. This is a relatively subtle effect~\cite{wilsonloops1, wilsonloops2, smear}.

It has been possible to precisely identify this effect only after the introduction of an additional dimensional parameter $s\ge 0$. $s$ is a resolution scale and is not physical in the sense that it is not measurable in any experiment. All elementary fields (gauge connections) now depend on, in addition to the location ${\vec x}$, one extra scalar length variable, $\sqrt{s}$. Keeping all $s$-arguments of a correlation function fixed and non-zero makes it singularity free for all the ${\vec x}$'s in a bounded region. For simplicity, one keeps all $s$ arguments non-zero and equal; setting $s=0$ brings us back to the
familiar distributional correlation functions.\footnote{Perhaps a better way to get back at un-smeared traditionally renormalized continuum correlation functions in Field Theory is to work with different values of the parameters $s$, $s_1,s_2\cdots$ going with each argument $x_1, x_2\cdots$ and take them sequentially to zero. One then needs a characterization of the limit that is invariant under a permutation of only the $s_i$ variables. This might be a way for the combinatorics of the standard OPE to emerge. }
 For $s>0$ the space-time correlation functions are, in particular, locally integrable, and the building of local composite operators is trivial. 

The operators at $s>0$ do not yet have any compelling applications to observable physics. Conceptually, they are useful in the context of YM because they highlight the crucial high resolution -- low resolution crossover occurring in a confining asymptotically free gauge theory. 

That the introduction of $s$ is well defined in the full theory cannot be proven, only checked numerically. The case is as strong as most other claims at this level. That continuous smearing works 
order by order in asymptotic perturbation theory is trivial conceptually (because ultraviolet divergences occur only in loop components of diagrams and not in tree components: this 
is why renormalization theory is typically applied to the generator of 1PI diagrams); explicitly working out the details seems to be, so far, somewhat tedious. 

Continuous smearing is defined by extending the YM fields to
$R^4\times R_+$ and by the equation $F_{\mu,s} = D_{\nu} F_{\mu,\nu}$ there. $F,D$ are the field strength and covariant derivative 
respectively and the equation is unique after minimizing the number of derivatives while requiring 5d gauge covariance and 4d rotational covariance in the $R^4$ component. As subscripts, $\mu,\nu$ are directions in the $R^4$ components and $s$ is the direction in the $R_+$ component. A boundary condition, $ A_\mu(x;s=0) = B_\mu (x)$, is attached to the equation. The boundary field $B$ is the ordinary quantum field on $R^4$. It is a random variable distributed according to the probability functional characterizing the Path Integral. The standard gauge choice is $A_s=0$ and produces a gradient-flow equation. The non-trivial point is that the ``rough'' character of the $x \in R^4$ dependence of $B_\mu(x)$ 
is eliminated after any finite amount of evolution in $s$, while the random character is maintained~\cite{wilsonloops1}.

At the classical level, the gradient flow has been used in analysis before. Among many examples, one is particularly relevant to my topic: In the
context of the PDEs of classical General Relativity it has been used as a way to devise a geometrically motivated Littlewood-Paley type decomposition on 2d compact manifolds which is generally covariant~\cite{seri}.
Clearly, in the 4d YM case something similar should take place. 

The continuum smearing equation in the partially gauge--fixed 5d form, $\partial_s A_\mu = D_{\nu} F_{\mu,\nu}$, retains  traditional 4d gauge covariance. Integrating this equation over $s$ in intervals $I_j=s_0 (1/2^{j+1},1/2^{j}], j\in Z$ produces a decomposition of $A_\mu (s=s_0/2^{j_1}) - A_\mu (s=s_0/2^{j_2})$ with $j_2 > j_1$ into pieces which transform homogeneously under gauge transformations since they are differences of gauge fields smeared by different amounts. The norms of the individual contributions corresponding to each $I_j$ segment are then gauge invariant. Increasing $j_2$ at fixed $j_1$ would allow to study the build-up of the gauge invariant piece of the expected UV singularities at $j_2=\infty$.

For the rest of the discussion in this paper I shall restrict myself to scalar fields with global symmetries where continuous smearing is simple, admitting a linear form. From Littlewood-Paley one naturally proceeds to more general wavelet decompositions. 

\section{Wavelets}

Our space of integration variables is the set of functions $\Phi$ from $R^d$ to $R$. This is a linear space and several norms can be defined on it. To make sense of the Path Integral one needs a countable basis so the integration is over the coefficients representing $\Phi$ in this basis. We also wish the order of the coefficients to be immaterial, so we are interested in unconditional bases. Thus, we are interested mainly in discrete wavelets. Constructions of wavelets typically start in $d=1$. Instead of $x\in R^d$ we shall use $t$ for $d=1$. 

The phase-space cell picture is related to standard coherent states, that is those associated with the group generated by $e^{i\alpha Q}, e^{i\beta P}, [Q,P]=i$. Coherent states generalized to other groups have been constructed by Perelomov~\cite{perelomov}. The states result by choosing one special state and acting on it with all the group elements. Typically, they span a well defined space, useful for Physics applications, but are over-complete. The problem of extracting a basis is addressed in~\cite{perelomov} in some examples.

The group associated with wavelets will be the group of affine  transformations $t\to at+b$ with $a\in R_+, b\in R$. Wavelets are defined in a manner similar to the construction of generalized coherent states: there is one standardized function, $\psi(t)$ on which the group acts. A discrete subset, consisting of $a=2^j$ and $b=n 2^j$, $n,j \in Z$, becomes an unconditional basis of $L_2$. The elements of the basis are 
$\psi_{j,k}(t)={\cal N}_j \psi (2^j t - k)$. ${\cal N}_j$ is a normalization, and $(j,k)\in Z^2$. 
$\int dt \psi(t) =0$ and $\psi_{j,k}(t)$ is concentrated in $t$-space around $t=2^{-j} k$ and ${\hat\psi}(\omega)$ around 
$\log_2 |\omega|= j$ ($\hat\psi (\omega)$ is the Fourier transform of $\psi(t)$). Each component along $\psi_{j,k}$ 
makes a contribution of resolving-power $~2^j$ around $2^{-j}k$.
The totality of $\psi_{j,k}(t)$ for fixed $j$ describes the
piece of $\Phi$ at resolution-power $~2^j$. 
As one scans index space from $j=-\infty$ \& all $k$ to $j=\infty$ \& all $k$ one goes from the deep IR to the deep UV. 

It is possible to find $\psi$'s such that the
$\psi_{jk}$ give unconditional orthonormal bases of the space of square integrable functions $f(t)$, $L_2$. The coefficients $c_{j,k}$ can be used to characterize also other spaces then $L_2$, like $L_p$. The mother wavelet $\psi$ can have some limited level of regularity, but is of a fractal nature when its support is compact, which is a much studied case. Obviously, the convergence of the expansion, which holds in $L_2$, cannot hold in
$L_1$ if $\int dt \Phi(t) \ne 0$. 

The limited support of $\psi(t)$ means that a local action $S[\Phi]$ will appear as a functional $S[c_{j,k}]$ which is 
local in $k$. A multinomial $S[\Phi]$ can be expanded into  $\sum_{j,k} F_1[c_{j,k}]+\sum_{(j,k)\ne (j',k')}
F_2[c_{j,k},c_{j',k'}]+\sum F_3[...]+\cdots$, and one wants to be dominated by the $F_n$ terms with very low $n$'s. Invariances of
the original action under translations of $\Phi$ are now 
invariances under shifts by $2^{-j}k$ (fixed $k$) 
of $c_{j,m}$ for all $m$. If in all $F_n$ the terms with 
any pair $(j,j')$, $|j-j'|\ge q$, where $q$ is some small integer, were negligible, one could 
define a transfer matrices ${\cal T}_j$ in the $-j$ direction (toward decreasing resolutions) such that the partition function $Z$ can be expressed as 
$Tr {\cal T}_{\infty}....{\cal T}_{j+1} {\cal T}_j{\cal T}_{j-1}....{\cal T}_{-\infty}$. This would be easy if the products stabilized on some ${\cal T}_{\pm\infty}$ at $j~\pm\infty$ which are bounded appropriately. This would produce an expression similar to the overlap used in lattice chirality. 
The history in $j$ would constitute an RG flow from a UV fixed point to an IR one, each corresponding to an extremal state of
${\cal T}_\pm$. Whether this works at all would depend on the choice for the ``mother wavelet'' $\psi$. The hope is that the
right choice would no longer seem like an act of ``fine-tuning''.

I have included a $j$ dependence on ${\cal T}$ because it may be necessary to make the couplings $j$-dependent. If we consider
a situation where there is no $j$-dependence we have both scale and translational symmetries. This would restrict the form of the
continuum action to one which obeys both symmetries. This may work in $d=1$, but may be too restrictive in higher dimensions.

The transfer matrices will be general matrices and there is no reflection positivity with the wavelets I am aware of. The order of indices is ``causal''. If we denote the spaces spanned by $\psi_{j,k\in Z}$ by $W_j$, we can define the subspace
$V_j=\cup_{j'\le j} W_{j'}$ and then $V_j \subset V_{j+1}$. Associated to $\psi$, the ``mother'' wavelet, there is 
a function $\varphi$, the ``father'' wavelet, whose translates span $V_j$. More precisely, $\varphi$ is associated with the conventional choice $j=0$ and the translates are by $k\in Z$, and $\varphi_{j',k'}$, is defined in terms of $\varphi$ in the same manner as $\psi_{j',k'}$ is in terms of $\psi$. Thus, the infinity at $j=-\infty$ can be eliminated and we can still span all of $L_2$ by adding 
$\varphi_{0,k\in Z}$ to the $j\ge 0, k\in Z$ set of $\psi_{j,k}$. This construction generalizes to $d>1$. 

For a well defined replacement of the partition function given by a Path Integral the set of index values of $(j\ge 0,k\in Z)$ must be made finite. The infinity of the range of $k$ is less problematic and might be avoided using the DLR framework. The necessary truncation left to do is in the UV, $j\le j_{UV}< \infty$.

Figuring out how to control $j$-infinite products of $j$-dependent 
transfer matrices may have some practical numerical merit in the context of constructing more stable actions for vector-like massless fermions~\cite{filters}.

Traditional Lattice Field Theory formulations have difficulty dealing with a complete RG trajectory connecting non-trivial 
RG fixed points. The actions generated during such a flow would likely explore the infinite space of actions in a much less controllable manner than in the case where one only wants to see the flow into, or away from, one fixed point~\cite{caswell}.

There is a vast literature on wavelets. I list some books in~\cite{waveletsbooks}, a sampling of some papers in~\cite{waveletspapers} and some papers with a Relativistic Field Theory orientation in~\cite{waveletsqft}. Of particular relevance might be~\cite{waveletstn} which shows analogies between tensor network RG approaches in Hamiltonian language and wavelets. I found~\cite{strang} particularly helpful to get a Physicist's grasp of wavelets. 

\subsection{Why Wavelets}

Wilson's analysis starts from the obvious observation that if the action can be written (I keep $d=1$ for simplicity) as $ S=\sum_{j,k} F_{j,k} (c_{j,k})$ the path integral factorizes and
one has to deal only with an infinite product. Therefore, one needs a basis which will provide as close a factorization as possible. At the least, one would want to be able to separate 
the action into a ``factorized'' form and an un-factorized correction in which one can expand in a controlled manner~\cite{wilson1, wilson2, battle}.  In the free field case we factorize the integral by Fourier transform, acting on functions in $L_2$. In a box the discrete Fourier coefficients make up a series in $l_2$ (space of countable sequences of real numbers whose squares add up to something finite). If a quartic term is present in the action, the fields should be also in $L_4$ and this is not simple to express in terms of Fourier coefficients. However, it would be simple in terms of wavelet coefficients: they would correspond to sequences also in $l_4$. In the interacting case the path integral no longer factorizes, but the un-factorizable piece can be expanded in because parts of the non-linearity are already included in the factorized part one is expanding about.

Wavelets were invented as tools in signal processing. The non-negative integrand of a certain Path Integral can be viewed as a black box generating signals with pre-determined relative probabilities. We sample the signal to estimate various quantities by statistical methods. Wavelets are useful because they can be adapted to the type of signals. To find good wavelets we need to characterize the main typical features of the signals.

In a gapless Quantum Field Theory there is a strong self-similarity in the momentum domain and also a strong local correlation in the relative probabilities between field configurations (our signals). Therefore we should be able to benefit from having a small set of functions on which dilations act simply. The actions of dilations on wavelets is linear and given by a matrix which is almost diagonal. In order to also represent translations in a simple manner the matrices representing dilations are best written in terms of
elementary shifts in all directions. We are doing this all the
time when we write down the kinetic term on some hypercubic lattice, and also think in these terms when imagining RG blocking procedures. 

\subsection{Wavelets, AdS/CFT, RQ}

Starting from a discrete infinite subgroup of the conformal group we obtained one segmentation of the path integral. Using wavelets we found another. Using RQ a third. 
The extra dimension in the two first cases measures the amount of relative scaling. The group behind the wavelet decomposition is a smaller subgroup of the conformal group than in the AdS/CFT case. In exchange it has translations, but no inversion, so differs substantially from RQ. 
The wavelet decomposition is less ambitious and sticks closer to translational invariance and to the traditional RG, where the flow is from high to low resolving power. This is why I think it is more workable.

\subsection{Wavelets for $d>1$.}

One generalization of wavelets to $d>1$ employs a basis made out of products of $d$ functions, one for each direction. For each direction $\mu$ one can pick either a $\varphi_{j,k_\mu} (x_\mu)$  
or a $\psi_{j,k_\mu}(x_\mu)$, except the case of all factors 
being $\varphi$'s is not allowed. For each overall resolution 
$j\in Z$ and each vector $k$ consisting of $d$-integers, one has $2^d-1$ basis elements. Let $\alpha=1,2,...(2^d-1)$ label the different elements. Notationally, one can represent $\alpha$ as a $d$-bit number $\alpha >0$. The set $\psi^\alpha_{j,k}$ is a basis of $L_2(R^d)$. The vicinity of $j=-\infty$ is again eliminated by extending the portion of this basis with $j\ge j_0$ with the states $\prod_{\mu=1}^d \varphi_{j_0, k_\mu} (x_\mu )$ 
$k\in Z^d$. All of $L_2$ is still spanned. 


The finite subgroup of rotations in $R^d$ that acts simply on the indices of the basis consists of permutations of the directions. 
One would need to alter the standard Daubechies wavelets if one
wished to extend the discrete rotation group to the full $d$-dimensional hypercube rotational symmetry group.

\section{Summary.}

The upshot is the following outline of a strategy to ameliorate fine-tuning of mass terms in a scalar field theory: Replace the integration variables $\Phi(x)$ in the Path Integral of a theory defined by an action $S[\Phi ]$ by the coefficients 
$\{c_{j\ge j_0,{\vec k}\in Z^d,\alpha }\}$ in a wavelet expansion $\Phi=\sum_{j\ge j_0,{ \vec k},\alpha} 
c_{j\ge j_0,{ \vec k},\alpha } \phi_{j\ge j_0,{ \vec k},\alpha}$, where $\alpha$ takes $2^d$ values for $j=j_0$ and $2^d -1$ values for $j_0 < j$. Start with an action that is classically scale invariant (for example $\lambda\Phi^6$ in $3d$). 

Dilations and translations act on $S[c]$ by acting on the set $\{c_{j\ge j_0,{\vec k}\in Z^d,\alpha }\}$ and, possibly on some couplings defining $S$ itself. Then, make some minimal needed changes in the action to get a new action $S'[c]$ such that one can conveniently quantize in the $j$-direction. By inverting the $\Phi\to c$ map check by how much locality of $S'[\Phi]$ is violated. The hope is that an acceptable violation will appear possible so that universal features at the fixed points are unaltered, but some fine tuning is eliminated. Maybe the index space labelling the wavelet basis ($\alpha$ might be absorbed into the fields' definitions) ``fuses'' in some sense into a sort of continuum of higher dimension where the rules determining allowed actions are simple and produce familiar gapless theories without fine-tuning.

\begin{acknowledgments}
This research was supported by the
NSF under award PHY-1415525. 
\end{acknowledgments}

\section{References Cited}

\clearpage

\end{document}